\documentstyle[psfig,subfigure]{mn}

\begin{document}
\title[Radio-Submillimetre Redshift Indicator]{Constraining the
Radio-Submillimetre Redshift Indicator using data from the SCUBA Local Universe Galaxy Survey.}
\author[L. Dunne, D.L. Clements, S. A. Eales]
{Loretta Dunne$^{1}$, David L. Clements$^{1}$, Stephen A. Eales$^1$\\
$^1$Department of Physics and Astronomy, University of Wales Cardiff,
PO Box 913, Cardiff, CF2 3YB\\ }

\maketitle

\begin{abstract}
Many of the faint submm sources uncovered by deep SCUBA surveys still
remain unidentified at most, if not all other wavelengths. The most
successful method so far of obtaining accurate positions and hence
allowing more secure identifications has been to use centimetre
wavelength imaging with the VLA. This has led to a tempting approach
for obtaining redshift estimates for this population (Carilli \& Yun
1999,2000), which relies on the tight FIR--radio relationship and
takes advantage of the steep spectral slope in the submm. In this
paper we use the submm data from the SCUBA Local Universe Galaxy
Survey (SLUGS) to estimate the usefulness of, and the uncertainties
in, the radio--submm redshift estimator. If the submm--radio spectral
index were correlated with either dust temperature or 850$\mu$m
luminosity, this method could produce biased redshift estimates for
850$\mu$m selected galaxies. We find, however, that within SLUGS,
these correlations are not significant. The ratio of 850$\mu$m/1.4 GHz
flux was found to decrease with increasing radio and FIR luminosity,
and we propose that this is due to a component of the dust not
associated with recent star formation, but which is instead heated to
15--20 K by the general interstellar radiation field.
\end{abstract}

\begin{keywords}
galaxies;high-redshift -- galaxies;submm -- galaxies;radio -- galaxies;primeval
\end{keywords}

\section{Introduction}
Ever since the discovery of Ultra-luminous Infrared Galaxies (ULIRGs)
(Joseph et al. 1985; Sanders et al. 1988) which emit up to 99 per cent
of their bolometric luminosity in the far-IR, there have been
suspicions that our optical/UV view of the early universe might be
biased by large amounts of dust obscuration.  These were supported by
the discovery of the Cosmic infrared Background (CIB) (Puget et
al. 1996; Fixsen et al. 1998), which contains up to twice the
luminosity of the integrated optical/UV background. It thus seems
likely that much of the star formation activity in the early universe
takes place in obscured environments. Many deep surveys have been
conducted with the SCUBA bolometer array on the James Clerk Maxwell
Telescope (JCMT)\footnote{The JCMT is operated by the Joint Astronomy
Center on behalf of the UK Particle Physics and Astronomy Research
Council, the Netherlands Organization for scientific Research and the
Canadian National Research Council.} in order to uncover the sources
responsible for this background. Significant numbers of sources, far
in excess of no evolution models have been found at both submm (Smail,
Ivison \& Blain 1997; Barger et al. 1998; Hughes et al. 1998; Eales et
al. 1999) and far-IR (Puget et al. 1999; Kawara et al. 1998)
wavelengths.

The next step after finding these sources is identifying them and
determining their redshifts, thus allowing the true star formation
history of the universe to be traced. This has proved to be a
difficult task, due to the large SCUBA beam size at 850$\mu$m (15
arcsec) and the several arcsec positional uncertainty inherent in all
of the SCUBA maps (Ivison et al. 1998,2000; Downes et al. 2000;
Barger, Cowie \& Richards 2000; Eales et al. 2000). The SCUBA sources
can sometimes be identified with several plausible optical candidates,
or in other cases they remain optical blank fields, even to the
magnitudes probed by HST and Keck. A possible solution to the problem
of poor positional accuracy lies in the tight correlation between
thermal FIR emission and synchrotron radio emission seen in the local
universe (Helou et al. 1985; Condon 1992). The mechanism for this
relationship is believed to be the massive stars from recent
star-formation, which both heat the dust to produce the FIR flux and,
when they explode as supernovae, provide relativistic electrons which
constitute the synchrotron radio emission. This means that radio
observations with the VLA are also sensitive probes of star formation,
albeit not as sensitive as the submm at higher redshifts given the
unfavourable K-correction at radio wavelengths. For example, Barger,
Cowie
\& Richards (2000) have recently targeted 1.4 GHz sources in the deep
VLA images (Richards 1999) of the Hubble Flanking Fields (HFF) with
SCUBA and found that the majority of bright ($> 6$ mJy) submm sources
have radio counterparts. By observing the submm sources from the deep
surveys at radio frequencies (particularly 1.4 GHz), more accurate
positions can be obtained, allowing more secure identifications of the
submm sources with optical or near IR counterparts. However, even with the
accurate VLA positions, many SCUBA sources still have no detectable
optical/IR counterparts, or they are too faint for spectroscopy. This
creates a great problem in determining the redshifts for the sources,
necessary to trace their evolution.

One possible route for obtaining redshift estimates for these objects
has been suggested by Carilli \& Yun (1999). This is based on the
FIR-radio correlation plus the break in the spectral slope at around 3
mm, where the thermal dust emission takes over from the declining
synchrotron tail. This puts the radio on one side of the spectral
break and the FIR (or submm in this case) on the other, creating a
redshift sensitive ratio. Carilli \& Yun (1999) used the submm-radio
spectral index, defined as:
\[ \alpha^{850}_{1.4} = 0.42 \times \log\left(\frac{S_{850}}{S_{1.4}}\right)
\]
where $S_{850}$ and $S_{1.4}$ are the fluxes at 850$\mu$m and 1.4
GHz.

The exact dependence of $\alpha^{850}_{1.4}$ on redshift is sensitive to
a few parameters:
\begin{enumerate}
\item{The slope of the thermal Rayleigh-Jeans tail in the FIR and
submm. This depends on frequency as $\nu^{2+\beta}$ where the dust
emissivity index ($\beta$) is thought to lie between 1 and 2 (Hildebrand
1983). It is the steepness of this slope which potentially makes the
indicator sensitive to redshift}
\item{The slope of the radio synchrotron emission. For star forming
regions this is assumed to be $-0.7\,\Rightarrow \,-0.8$ (Condon 1992). At
higher frequencies this will flatten due to radio free-free emission
but this is not simple to model. Free-free absorption at lower
frequencies can also lead to a flattening of the radio spectra, the
likely effect of this on $\alpha^{850}_{1.4}$ is discussed in Carilli
\& Yun (2000) and they conclude that it is unlikely to be a dominant
cause of scatter in the relationship. In any case the slope at radio
frequencies is much shallower than in the submm and any uncertainties
in it have less effect.}
\item{The temperature of the dust. This determines the redshift at
which the thermal spectrum turns over. When this happens the ratio
only depends weakly on redshift, and so becomes useless as a redshift
indicator. Accurate dust temperatures are notoriously difficult to
estimate and require measurements at many different wavelengths
throughout the submm and FIR.}
\item{Contribution by AGN. Galaxies containing a radio-loud AGN will
have higher radio emission than a purely star-forming galaxy and so
lower values of $\alpha^{850}_{1.4}$. This could lead to ambiguity
between high redshift objects with AGN and lower redshift starbursts.} 
\item{The FIR-radio correlation and its variability with redshift. In
using this technique to estimate redshifts, an inherent assumption is
made that the FIR-radio correlation does not vary with
redshift. Possible causes of such a change could be a variation in
magnetic field strengths in galaxies in the past, or changes in the
dust mass opacity coefficient which relates the FIR/submm emission to
the mass of dust. The latter would depend on the dust composition and
grain sizes so there is scope for it to have been different at the
epoch of galaxy formation, when metallicities would have been lower.}
\item{Linked to (v) is the possibility that the FIR-radio, or more
importantly, the submm-radio relationship itself depends on some other
galaxy property such as dust temperature, luminosity etc.}
\end{enumerate}
The possible impact of points i--vi above will be discussed later.

Several authors have already used this technique to determine redshift
estimates or limits for sources discovered in deep submm surveys with
SCUBA (Hughes et al. 1998; Lilly et al. 1999; Smail et al. 1999;
Barger et al. 2000; Eales et al. 2000). This paper uses data from the
SCUBA Local Universe Galaxy Survey (SLUGS) to examine the usefulness
of the submm-radio ratio as a redshift indicator, to try to normalise
it with respect to the local universe and to estimate the
uncertainties in it.

We will use $H_0=75$ km s$^{-1}$ Mpc$^{-1}$ and $q_0 = 0.5$ throughout.

\section{Constraining the Redshift Indicator}

The low redshift data is taken from the SLUGS survey (Dunne et
al. 2000) which is a complete sample of 104 galaxies selected from the
{\em IRAS\/} Bright Galaxy Sample. The sample includes mainly
infra-red bright objects ($L_{\rm{fir}} = 7 \times 10^9 - 2 \times 10^{12}\, 
\rm{L_{\odot}}$), some ULIRGS as well as many `more normal'
galaxies. The objects were mapped at 850$\mu$m with SCUBA and so
should provide reliable submillimetre fluxes. We fitted a
single-temperature dust spectral energy distribution (SED) to the
measured fluxes for each galaxy, allowing the dust emissivity index
($\beta$) as well as the temperature to vary to find the best
fit. While the assumption of a single temperature is probably wrong,
and indeed the low values of $\beta$ we derive suggest that there is a
cold dust component, the SEDs are still a good empirical fit to the
data, which is all that we require for the present analysis. 

The original Carilli-Yun models for the behaviour of the radio-submm
spectral index with redshift were simple two power-law models, along
with some empirical SED data for 2 local galaxies. They have since
modified their original estimator by using data from nearby galaxies (17 
objects) to try to provide a normalisation and to estimate the scatter 
in the submm-radio ratio with redshift (Carilli \& Yun 2000). We have
independently undertaken a similar analysis but using the full SLUGS
data set (104 objects), thereby providing a more statistically valid
measure of the relationship and its scatter.

\begin{figure}
\vspace{11cm}
\includegraphics{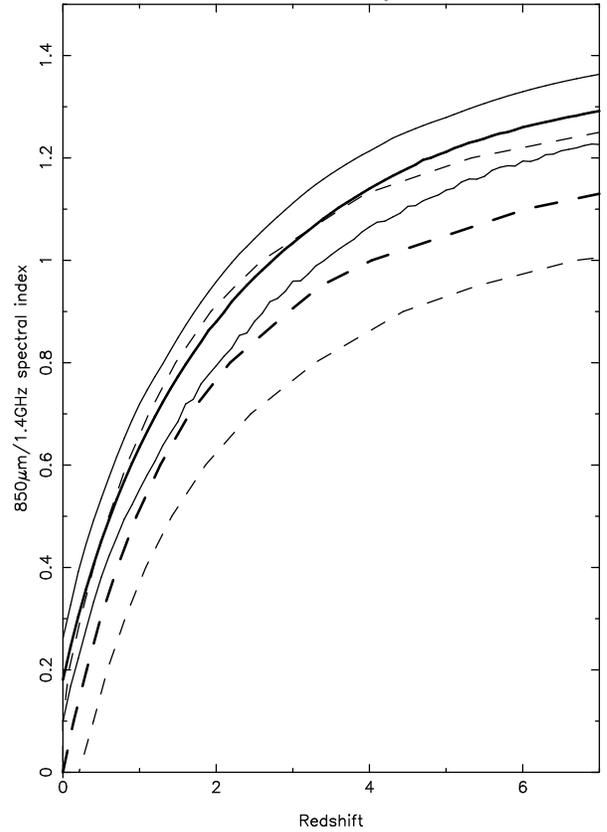}
\caption{\label{carF} The median (thick solid) and $\pm \,1 \sigma$ (thin
solid) estimators using the SLUGS low redshift data. The recent
Carilli \& Yun (2000) estimator is shown (dashed) with $\pm 1 \sigma$
uncertainties for comparison. The differences in redshift estimates
between this work and the Carilli-Yun estimator at various values of
$\alpha^{850}_{1.4}$ are given in Table~\ref{zcompT}.}
\end{figure}

We have taken the real spectral energy distributions of the 104 SLUGS
galaxies as determined from the {\em IRAS\/} 60 and 100$\mu$m fluxes
and the SCUBA 850$\mu$m data, and the 1.4 GHz radio flux (Condon et
al. 1990). The measured radio spectral indices were used where
available (for about half of the sample) and the median
$\alpha_{radio}$ was found to be $-0.7$. This was used in the SEDs of
those galaxies with no measured $\alpha_{radio}$. If the change in
each of these SEDs with redshift is plotted, we
get 104 separate curves from which the median can be selected. The
uncertainty in this method (at any given redshift) can be estimated
from the spread in the 104 curves. In Figure~\ref{carF}, the thick
solid line shows the median curve with the upper and lower solid
curves being those ranked 16 per cent from the top and bottom
respectively, to give the $\pm 1 \sigma$ errors (defined as the range
of $\alpha^{850}_{1.4}$ in which 68 per cent of the lines lie). So at
any given redshift, the horizontal distance between the thin lines
provides an estimate of the $\pm 1 \sigma$ uncertainty on the redshift
provided by the median curve. The rms scatter in $\alpha^{850}_{1.4}$
is almost constant with redshift, decreasing very slightly from 0.097
at $z=0$ to 0.080 at $z=6$. A rough guide to the redshift
uncertainties at various redshifts are given in
Table~\ref{errorT}. The median curve can be approximated by the
expression
\[z=0.551-6.652\alpha+25.57\alpha^2-30.56\alpha^3+13.75\alpha^4
\]
Our estimator starts to loose its effectiveness at
redshifts greater than about 4--5, where it starts to turn over due to
the peak in the thermal dust spectrum. This can be seen from the
increase in $\sigma_z$ with redshift (Table~\ref{errorT}). Also, as
mentioned in Carilli \& Yun (1999) at $z > 6$ the radio emission
begins to become quenched by inverse Compton losses off the microwave
background. The dashed line in Fig.~\ref{carF} shows the new
Carilli-Yun estimator (with $\pm 1\sigma$ uncertainties) using the
data from 17 local galaxies (Carilli
\& Yun 2000). It can be seen that their mean estimator lies outside
the 1$\sigma$ uncertainties from our analysis, however the upper $1
\sigma$ curve from theirs is consistent with our lower $1 \sigma$
limit, with agreement being better at lower redshifts ($<2$). The
differences in redshift estimates from the two indicators are
listed in Table~\ref{zcompT} for various values of $\alpha^{850}_{1.4}$.

\begin{table}
\caption{\label{errorT} $1 \sigma$ uncertainties on $Z_{med}$}
\begin{tabular}{cc}
\multicolumn{1}{c}{$Z_{med}$}&\multicolumn{1}{c}{$\pm \sigma_{z}$}\\
\\[-1ex]
1.0 & 0.28\\
2.0 & 0.44\\
3.0 & 0.63\\
4.0 & 0.94\\
5.0 & 1.25\\
6.0 & $+$3.13\\
    & $-$1.44\\
\hline
\multicolumn{2}{p{26em}}{\normalsize {Uncertainties based on scatter in low
redshift data only, there will be additional uncertainty when the
flux errors of the high-z sources are accounted for.}}\\
\end{tabular}
\end{table}

There are two main differences between the present results
and those in Carilli \& Yun (2000). The first is in the zero redshift
normalisation in the two samples. The mean value of
$\alpha^{850}_{1.4}$ at $z=0$ in Carilli \& Yun (2000) is $\approx 0$
with an rms scatter of 0.14, while for our sample the mean
$\alpha^{850}_{1.4}= 0.18$ at $z=0$, with a lower rms scatter of
0.097. The shape of the Carilli-Yun estimator is also different due to
the SED shapes of the galaxies in their sample (which in
general had higher values of $\beta$ than ours). These differences are
partly due to our much larger sample size, but there are also
discrepancies in the submillimetre fluxes of the individual galaxies
which are present in both the samples used by ourselves and Carilli \&
Yun (2000). Most of their submm data comes from a sample of 19 bright
{\em IRAS\/} galaxies (Lisenfeld, Isaak \& Hills 2000). Seven of the
galaxies in this sample were also observed by us, and in several cases
the fluxes we derived were as much as a factor of two higher. This was
attributed to Lisenfeld et al. missing flux from extended objects
(Lisenfeld, private communication), and has been corrected in the
revised version of the Lisenfeld et al. paper. The consequences of
using under-estimated submm fluxes would be to i) lower the mean
$\alpha^{850}_{1.4}$ and ii) increase the value of $\beta$ determined
in single temperature fits to the 60, 100 and 850$\mu$m fluxes, which
explains the different shape of the Carilli-Yun estimator. We also
note a possible error in the 1.4 GHz fluxes used for Zw049.057 and NGC
5936 by Carilli \& Yun (2000). An erratum has since been published to
accompany Carilli \& Yun (2000) and this is discussed further in
Section 5.

\begin{figure}
\vspace{11cm}
\includegraphics{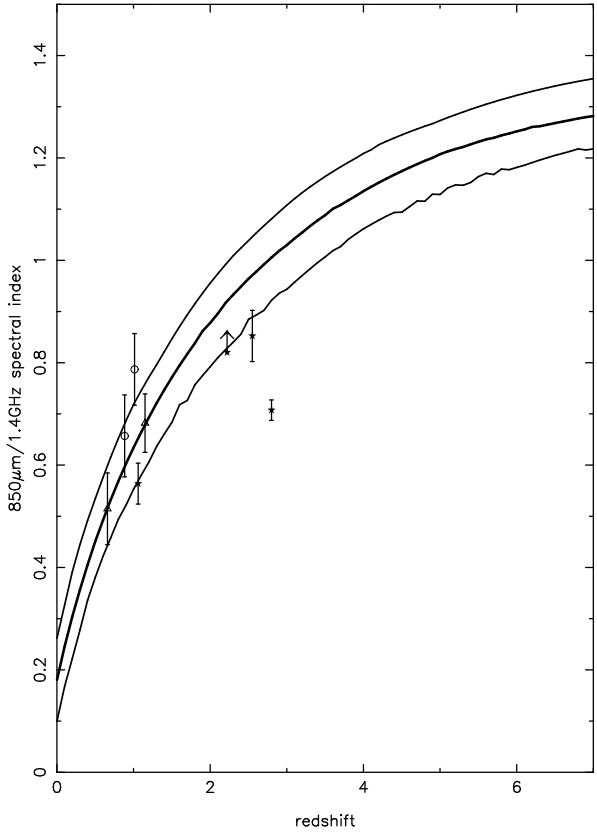}
\caption{\label{car2F} The median (thick solid) and $\pm \,1 \sigma$ (thin
solid) estimators using the SLUGS low redshift data. Data points are
sources from the deep submm surveys which have secure radio fluxes and
spectroscopic redshifts. Triangles are from the Canada-UK deep survey
(Eales et al. 1999; Lilly et al. 1999; Eales et al. 2000). The point
at $z=1.15$ is from Eales et al. 2000 and has fluxes of
$S_{850}=3.3$mJy, $S_{5\rm{GHz}}=53.6\mu$Jy with
$\alpha_{radio}=-0.3$. Stars are from the Cluster survey (Ivison et
al. 1998; Smail et al. 2000; Ivison et al. 2000). This includes a
$z=2.22$ galaxy which has a radio upper limit. Circles are from the
SCUBA observations of radio sources in the HFF (Barger et
al. 2000). The point at $z=2.8$, which lies well below all the curves,
is SMM02399-0136 from Ivison et al. (1998) and is believed to be an
AGN.}
\end{figure} 

\begin{table}
\caption{\label{zcompT} A comparison of estimated median redshifts and 
ranges of redshift for the relationship in this work and that of
Carill \& Yun (2000)}
\begin{tabular}{ccccc}
\multicolumn{1}{c}{$\alpha^{850}_{1.4}$}&\multicolumn{1}{c}{$Z_{med}$}&\multicolumn{1}{c}{$Z_{med}$}&\multicolumn{1}{c}{$Z^{+\sigma} -
Z^{-\sigma}$}&\multicolumn{1}{c}{$Z^{+\sigma} - Z^{-\sigma}$}\\
\multicolumn{1}{c}{}&\multicolumn{1}{c}{SLUGS}&\multicolumn{1}{c}{CY}&\multicolumn{1}{c}{SLUGS}&\multicolumn{1}{c}{CY}\\
\\[-1ex]
0.3 & 0.19 & 0.49 & 0.07--0.34 & 0.23--0.81\\
0.5 & 0.64 & 0.96 & 0.42--0.82 & 0.60--1.42\\
0.7 & 1.20 & 1.65 & 0.93--1.54 & 1.12--2.45\\
0.9 & 2.12 & 2.95 & 1.73--2.62 & 1.93--4.44\\
1.1 & 3.57 & 6.04 & 2.92--4.46 & 3.63--$>7$\\
\hline
\multicolumn{5}{p{26em}}{\normalsize{\em Column(1)}-Value of spectral
index; {\em Column(2)}-redshift estimate using median line from SLUGS; 
{\em Column(3)}-as for (2) but using Carilli \& Yun (2000) estimator;
{\em column(4)}-range of redshifts between $\pm\,1\sigma$ curves from
SLUGS; {\em Column(5)}-as for (4) but using CY (2000).}\\
\end{tabular}
\end{table}  

If we now place the handful of SCUBA sources from the deep surveys
which have both spectroscopic redshifts and radio fluxes (Smail et
al. 1999 and references therein; Eales et al. 1999; Lilly et al. 1999;
Barger et al. 2000; Eales et al. 2000; Ivison et al. 2000) on the
redshift estimator we have created from our observed SEDs
(Fig.~\ref{car2F}), we see that the agreement is satisfactory given
the uncertainties. The object at $z=2.8$ is SMM02399-0136 from Ivison
et al. (1998) and is known to harbour an AGN, which explains why it
lies below the predicted lines. Given that sources may have AGN
activity but cannot be confirmed as such, this redshift estimator can
give only reliable lower limits. Additionally, if taken with another
redshift estimation, such as photometric redshift, it could help to
identify the presence of AGN (a point first noted by Carilli \& Yun
1999).

\renewcommand{\subfigcapskip}{2pt}
\renewcommand{\subfigcapmargin}{20pt}

\newcommand{\goodgap}{%
 \hspace{\subfigtopskip}%
 \hspace{\subfigbottomskip}}

\begin{figure*}
  \subfigure[FIR--radio correlation]{\psfig{file=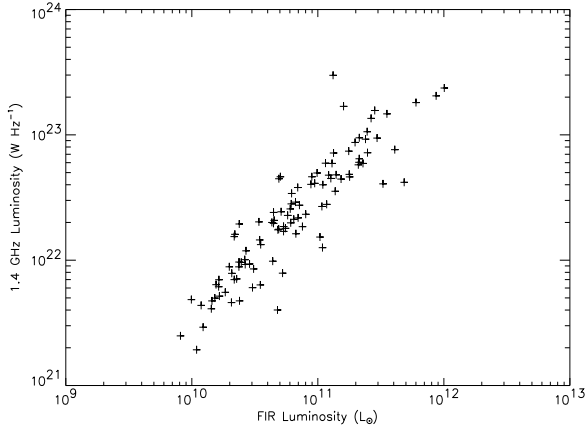,width=8cm,height=6cm}}\goodgap
  \subfigure[850$\mu$m--radio correlation]{\psfig{file=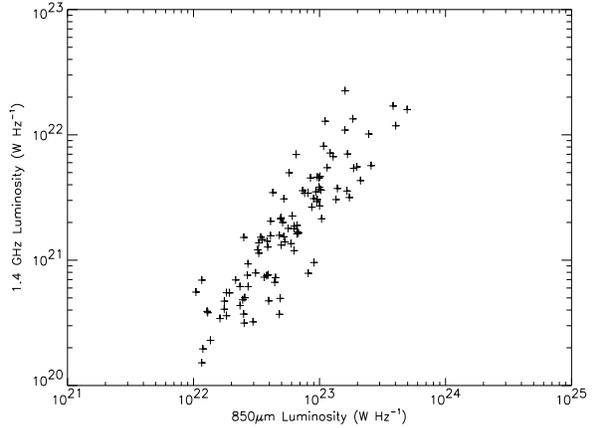,width=8cm,height=6cm}}\\
\caption{\label{firradF} The FIR--radio relation and the
850$\mu$m--radio relation for the SLUGS galaxies. The FIR--radio
correlation is the tighter one implying that this is where the
physical connection lies.}
\end{figure*}

\section{The FIR and Submm Vs. Radio Properties}

\begin{figure*}
  \subfigure[Dependence of $\alpha^{850}_{1.4}$ on $L_{\rm{fir}}$,
first noted by Carilli \& Yun (2000).]{\psfig{file=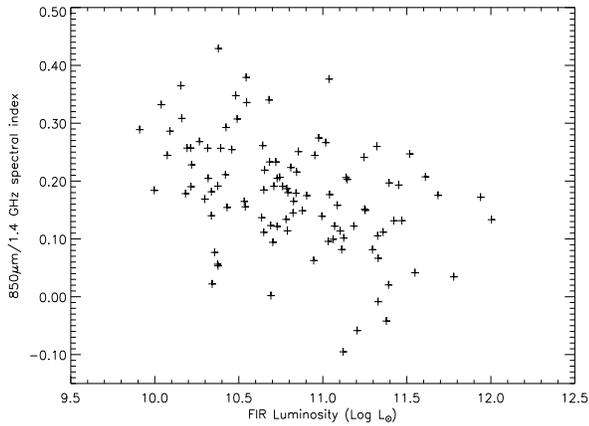,width=8cm,height=6cm}}\goodgap
  \subfigure[A stronger correlation is seen with $L_{1.4}$]{\psfig{file=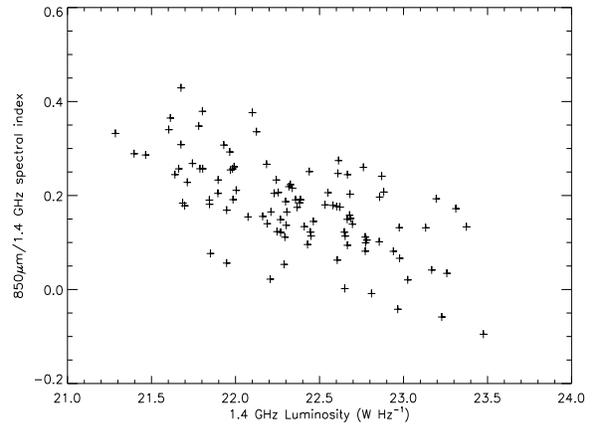,width=8cm,height=6cm}}\\
  \subfigure[$\alpha^{850}_{1.4}$ shows no significant correlation with $L_{850}$]{\psfig{file=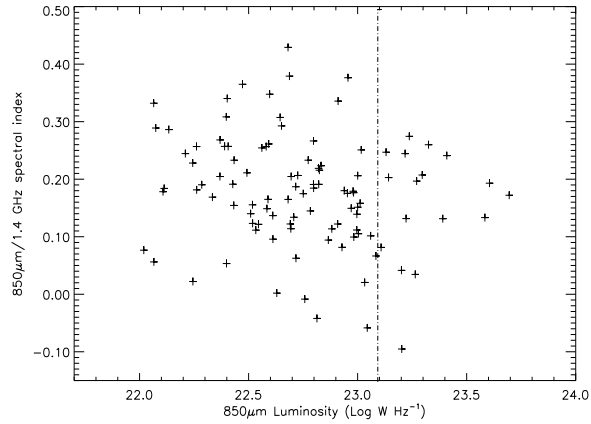,width=8cm,height=6cm}}\\
\caption{\label{alpha_lumF} Relationships between $\alpha^{850}_{1.4}$ 
and luminosities for the SLUGS galaxies. The strongest connection is
with radio luminosity while there is no strong dependence on
$L_{850}$. The dot-dash line in (c) shows a rough lower limit to the
rest-frame 850$\mu$m luminosity of a deep SCUBA source with
$S_{850}=4$mJy, $z>2$, $\beta=2$ and $T_{\rm{d}}=50$ K. Lower $\beta$
or $T_{\rm{d}}$ will produce slightly higher luminosities. The flat
flux density-redshift behaviour at 850$\mu$m means that sources
brighter than 4mJy will have $L_{850}$ higher than the line, and
vice-versa for fainter ones.}
\end{figure*}

The SLUGS sample provides an ideal test for the variation of the
FIR-radio and submm-radio relationships with galaxy properties, such
as dust temperatures, luminosities etc. which may affect the
reliability of $\alpha^{850}_{1.4}$ as a redshift indicator, and which
is also interesting in its own right. Figure~\ref{firradF} shows the
FIR-radio correlation for the SLUGS galaxies along with the
850$\mu$m--1.4 GHz relationship. Both are highly significant
(Table~\ref{corrT} lists all the correlation coefficients in this
discussion) but the scatter in the FIR-radio correlation is smaller
(0.06 (15\%)) than that which exists for the submm-radio (0.097
(25\%)). This suggests that it is actually the FIR which is physically
related to the radio rather than the submm. There is also a difference
in slope between the two, which we will return to later. The FIR-radio
correlation in the local universe does not depend on galaxy luminosity
(except at very low $L_{\rm{fir}}$) or dust temperature, something
which is confirmed by this sample. However if the submm-radio
relationship is tested in this way by plotting $\alpha^{850}_{1.4}$
against $L_{\rm{fir}}$, $L_{1.4}$ and $L_{850}$
(Figure~\ref{alpha_lumF}) then it is clear that there is a strong
dependence of $\alpha^{850}_{1.4}$ on both $L_{\rm{fir}}$ and
$L_{1.4}$ but not on $L_{850}$. The link with $L_{\rm{fir}}$ was noted
by Carilli \& Yun (2000) but the trend is even more evident with radio
luminosity. The most likely explanation for these correlations is that
both $L_{\rm{fir}}$ and $L_{1.4}$ are very sensitive to recent star
formation in a galaxy but $L_{850}$ is not as sensitive, as a
significant fraction of the 850$\mu$m flux may be produced by colder
dust heated by the general interstellar radiation field
(ISRF). Therefore $L_{850}$ will not change by as great a fraction as
$L_{\rm{fir}}$ with increasing star formation. This explains the
larger scatter and steeper slope in the 850$\mu$m--1.4 GHz
correlation. 

One of the two crucial relations for determining whether
$\alpha^{850}_{1.4}$ is a biased redshift indicator is that between
$\alpha^{850}_{1.4}$ and $L_{850}$. Since SLUGS was selected at
60$\mu$m, it is quite possible that the 850$\mu$m luminosities of
SCUBA-selected galaxies are distinctly different from those in
SLUGS. If there were also a significant relationship between
$\alpha^{850}_{1.4}$ and $L_{850}$, the calibration of the Carilli-Yun
method using SLUGS galaxies would lead to biased redshift
estimates. Fig.~\ref{alpha_lumF}(c) shows, however, that there is no
significant correlation. To examine this, the dot-dash line on
Fig.~\ref{alpha_lumF}(c) shows the lower end of the rest-frame
850$\mu$m luminosity range for a galaxy with $S_{850}=4$mJy at
redshifts of $>2$ ($\beta=2$ and $T_{\rm{d}}=50$ K, lower $\beta$ and
$T_{\rm{d}}$ give slightly higher rest-frame $L_{850}$ for the
same flux and redshift). If we look at the values of
$\alpha^{850}_{1.4}$ for the SLUGS galaxies, which correspond to the
luminosities of the high redshift objects (i.e. the crosses to the
right of the dot-dash line in Fig.~\ref{alpha_lumF}(c)), they do not
display a lower mean value than that for the whole sample. Given how
little we know about the properties of the high-z SCUBA galaxies, and
whether they are related to galaxies of the same luminosity at low
redshift (density evolution) or to galaxies of lower luminosity
(luminosity evolution), we do not feel that there is a need for a
luminosity dependent indicator. This is further supported by
Fig.~\ref{car2F}, showing a relatively good agreement of the indicator
with the high-$z$ data, and in particular, there is no systematic bias
from the line in one direction, which may be indicative of changes in
$\alpha^{850}_{1.4}$ with redshift/luminosity.

\begin{figure*}
  \subfigure[\label{alpha_tempF}No significant correlation is found
between $\alpha^{850}_{1.4}$ and the fitted dust temperature]{\psfig{file=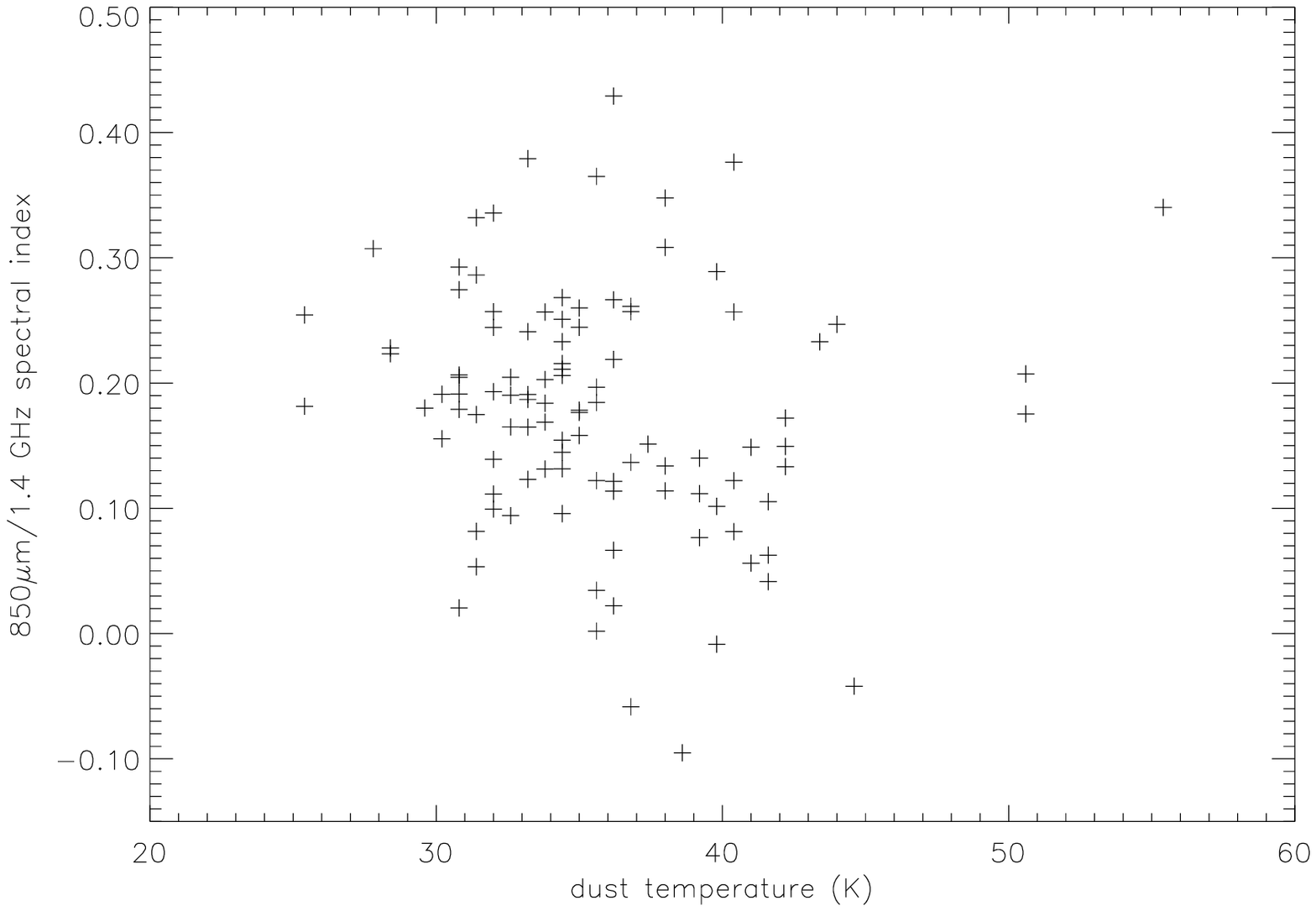,width=8cm,height=6cm}}\goodgap
  \subfigure[\label{alpha_betaF}Possible slight dependency of
$\alpha^{850}_{1.4}$ on emissivity index $\beta$, see text for
discussion]{\psfig{file=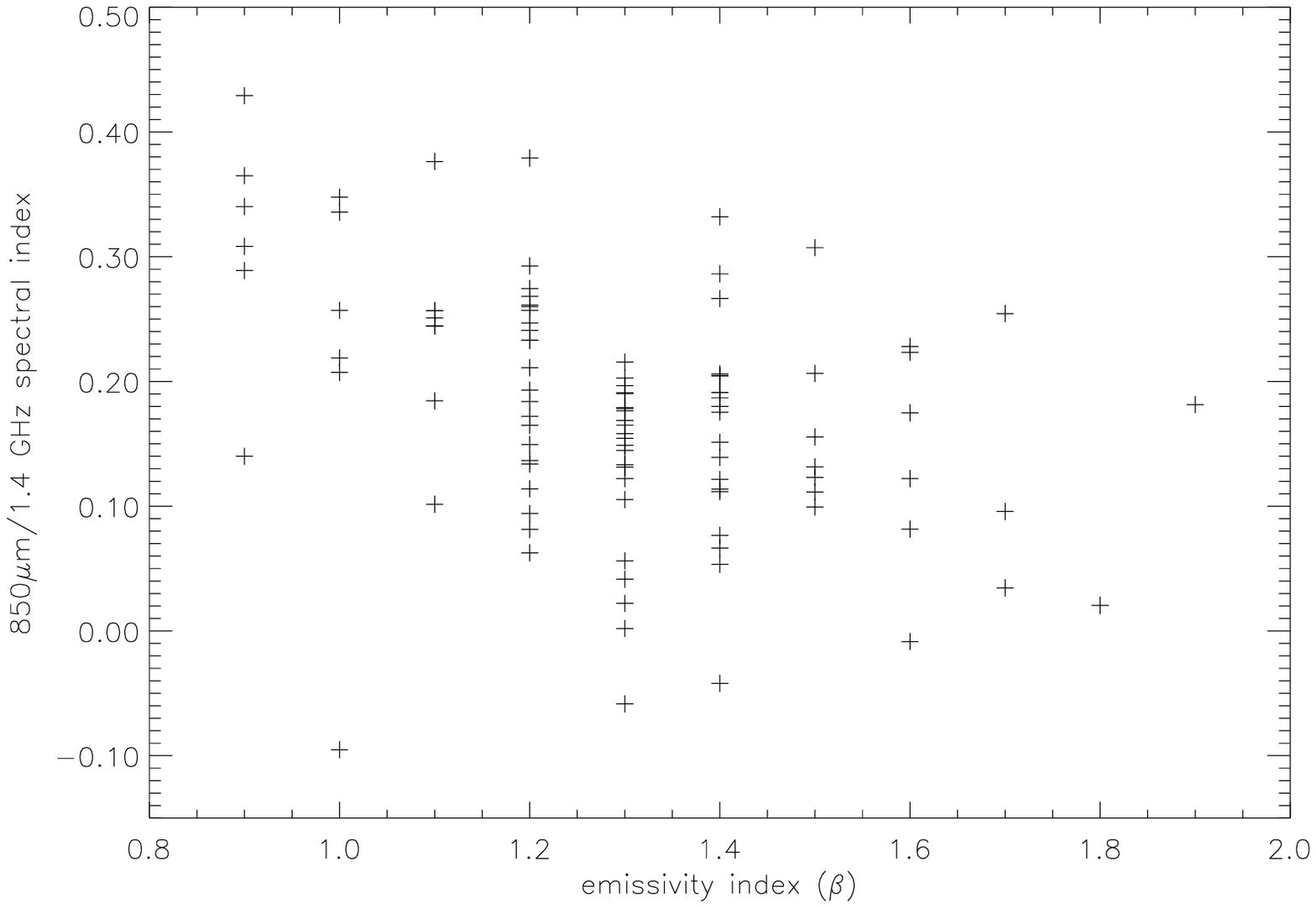,width=8cm,height=6cm}}\\
\caption{Variation of $\alpha^{850}_{1.4}$ with fitted dust
temperature and emissivity index}
\end{figure*}

The second important relationship is that between $\alpha^{850}_{1.4}$
and dust temperature, since an 850$\mu$m selected sample may well have
a different mean dust temperature than one selected at 60$\mu$m. It
has also been suggested by Blain (1999) that there should be a
dependence of $\alpha^{850}_{1.4}$ on dust temperature, and that this
produces a degeneracy in the redshift indicator whereby a hot galaxy
at high redshift may be indistinguishable from a colder galaxy at
lower redshift. We have tested for a temperature dependence by
plotting $\alpha^{850}_{1.4}$ against dust temperature, and we find no
significant correlation (Figure~\ref{alpha_tempF}). The slight
correlation seen between $\alpha^{850}_{1.4}$ and dust emissivity
index $\beta$ (Figure~\ref{alpha_betaF}) is probably linked to the
relationship of $\alpha^{850}_{1.4}$ with $L_{1.4}$ and
$L_{\rm{fir}}$. Despite the significant correlation given by the
statistic to this relationship, it appears to only hold for the lower
values of $\beta$ (0.9--1.4). This could be because a low $\beta$
(when produced by fits to 60,100 and 850$\mu$m points only) is another
possible indicator of the fraction of 850$\mu$m emission produced by
the ISRF, rather than directly by star forming regions. Evidence for
this comes from observations of a sample of optically selected
galaxies presently under way with SCUBA and also from studies of NGC
891 and the Milky Way (Alton et al. 1998; Masi et al. 1995), where the
SEDs using only 60, 100 and 850$\mu$m fluxes give a very low $\beta$
of $\sim 0.7$. These galaxies are known to have large fractions of
cold dust at $T < 20$ K, however this does not lead to a dependence of
$\alpha^{850}_{1.4}$ on the fitted dust temperature because that is
determined by the FIR fluxes which are dominated by the warmer dust.

\begin{table}
\caption{\label{corrT} Correlation coefficients.}
\begin{tabular}{ccccc}
\multicolumn{1}{c}{$y$}&\multicolumn{1}{c}{$x$}&\multicolumn{1}{c}{$r_s$}&\multicolumn{1}{c}{significance}&\multicolumn{1}{c}{S.D.}\\
\\[-1.5ex]
$L_{1.4}$ & $L_{\rm{fir}}$ & 0.92 & $1.5\times10^{-20}$ & $9 \sigma$ \\
\\[0.01pt]
$L_{1.4}$ & $L_{850}$ & 0.89 & $1.1\times10^{-19}$ & $9 \sigma$ \\
\\[0.01pt]
$\alpha^{850}_{1.4}$ & $L_{\rm{fir}}$ & $-0.43$ & $1.3\times10^{-5}$ & 
$4 \sigma$\\
\\[0.01pt]
$\alpha^{850}_{1.4}$ & $L_{1.4}$ & $-0.61$ & $5.4\times10^{-10}$ & $6 \sigma$\\
\\[0.01pt]
$\alpha^{850}_{1.4}$ & $L_{850}$ & $-0.22$ & 0.03 & $2 \sigma$\\
\\[0.01pt]
$\alpha^{850}_{1.4}$ & $T_{\rm{dust}}$ & $-0.22$ & 0.03 & $2 \sigma$\\
\\[0.01pt]
$\alpha^{850}_{1.4}$ & $\beta$ & $-0.42$ & $2\times10^{-5}$ & $4 \sigma$\\
\hline
\multicolumn{5}{p{27em}}{\sc notes-- \normalsize{{\em Column(3)}-Spearman
rank correlation coefficient; {\em Column(4)}-probability that $x$ and
$y$ are unrelated (i.e. of accepting the null hypothesis); {\em
Column(5)}-standard deviation at which the null hypothesis is
rejected.}}
\end{tabular}
\end{table}

\section{Discussion}

The submm-radio spectral index appears to provide a moderately
satisfactory redshift indicator out to redshifts of $\sim 4-5$ where
it ceases to be as sensitive to redshift. The uncertainties in the
estimated redshifts are obtainable using the spread of the local SEDs,
and range from $<\pm 0.3$ at $z<1$ to $>\pm 1$ at $z>4$.  Sources
which harbour radio loud AGN will not follow this pattern as they will
lie below the indicators, leading to AGN sources which are not recognised
as such being given misleadingly low redshifts.

\begin{table}
\caption{\label{sedT} Parameters used in `grey' cold component models.}
\begin{tabular}{lcccc}
\multicolumn{1}{c}{Model}&\multicolumn{1}{c}{$T_w$
(K)}&\multicolumn{1}{c}{$T_c$
(K)}&\multicolumn{1}{c}{$N_c/N_w$}&\multicolumn{1}{c}{$\beta$}\\%
\\[-1ex]
1 (thick dashed) & 40 & 20 & 5 & 2\\
2 (thin dashed) & 40 & 20 & 10 & 2\\
3 (thin dotted) & 40 & 20 & 40 & 2\\
4 (thick dotted) & 30 & 15 & 10 & 2\\
\hline
\multicolumn{5}{p{27em}}{\normalsize {{\em Column(1)}-Refers to model in
Fig~\ref{greyF}; {\em Column(2)}-Temperature of warm dust component;
{\em Column(3)}-Temperature of cold dust component; {\em
Column(4)}-Ratio of cold dust mass to warm; {\em Column(5)}-Value of
emissivity index assumed}}\\
\end{tabular}
\end{table}

\begin{figure}
\vspace{9.8cm}
\includegraphics{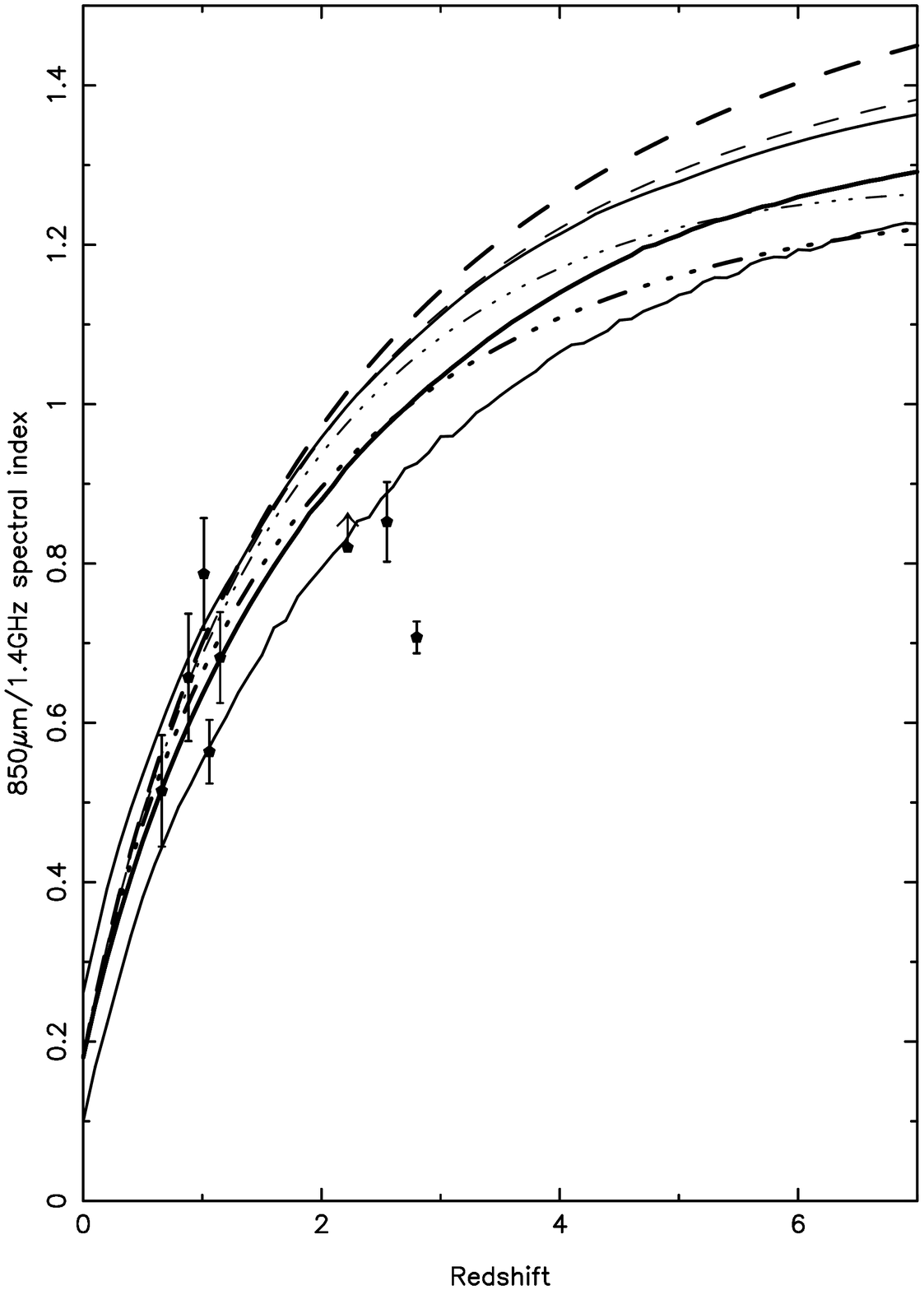}
\caption{\label{greyF} Median and $1\sigma$ estimators (solid lines)
along with models for two-component dust SEDs. Parameters are detailed
in Table~\ref{sedT}: $T_w =40$ K, $T_c=20$ K, $N_c/N_w = 5$ (thick
dashed); $T_w=40$ K, $T_c=20$ K, $N_c/N_w =10$ (thin dashed); $T_w=40$
K, $T_c=20$ K, $N_c/N_w = 40$ (thin dash-dotted); $T_w=30$ K, $T_c=15$ K,
$N_c/N_w=10$ (thick dash-dotted).}
\end{figure}

Let us now return to the assumptions and limitations discussed in
section 1, and apply them to our data sets.
\begin{enumerate}
\item{The submillimetre emissivity index has a mean value of 1.3 for
the SLUGS sample, as determined by single temperature fits to the 60,
100 and 850$\mu$m fluxes. As discussed at length in Dunne et
al. (2000), if a colder dust component is present this may not
represent the true value of $\beta$, the true value being higher than
that determined using a single component model. If the true value of
$\beta$ were nearer to 2 this would increase the rate of change of the
ratio with redshift as the submm flux in the Rayleigh-Jeans tail
$\propto \nu^{2+\beta}$ (see (iii)).}
\item{The radio spectral index is also somewhat uncertain, but being
flatter it has much less impact than the submm slope. However, if
high-redshift dust sources have systematically different spectral
indices than those at low redshift, this would lead to a bias. The
range of measured indices for the SLUGS sample is $\sim -0.2 $ to
$-1.0$. The situation at high redshifts is harder to determine as the
value of $\alpha_{radio}$ for $\mu$Jy sources may depend on the
selection frequency, with sources selected at $\nu>5$ GHz likely to
have flatter radio spectra than those selected at 1.4 GHz (Richards
1999). For example, the sources from the Canada-UK SCUBA survey have 5
GHz radio fluxes from Fomalont (1991) and an average $\alpha_{radio}
=-0.38$, while SCUBA observations of 1.4 GHz sources in the Hubble
Flanking Fields were associated with radio sources with steeper
spectra, $\alpha_{rad} \sim -0.7$ (Barger et al. 2000). This is most
probably only a selection effect, as only relatively flat spectrum
sources would be detected at 5 GHz. In general, it is preferable to survey 
at 1.4 GHz as this gives the best sensitivity to high redshift objects.}
\item{The mean dust temperature in the local galaxies could be
over-estimated. If a colder component is present then it will produce
a flattening of the thermal dust spectrum at longer wavelengths when
compared to warm dust, thus lowering the redshift at which the
estimator loses its sensitivity to redshift. We can model the
implications of points (i) and (iii) as they are inter-related. If there
is a cold dust component in the local galaxies, the true value of
$\beta$ is likely to be nearer 2 and the mean dust temperature
lower. If we make assumptions about the temperatures of the two
components and the relative masses in each (and that $\beta=2$), we
can then produce a new indicator. Figure~\ref{greyF} shows these
`grey' indicators for a few values of the various parameters which
reflect current observational evidence (Frayer et al. 1999; Alton et
al. 1998; Calzetti et al. 2000). The parameters used are given in
Table~\ref{sedT}, and all `grey' models were normalised at zero
redshift to have the mean value for the sample (0.18). There is
generally no great difference until $z
\sim 2$, and at this point the different parameters assumed start to
make a larger impact. In general, we might expect that the galaxies
detected by both SCUBA and the VLA would have $T_w > 30$ K, since it
is the recent star formation which the VLA is sensitive to (and which
also produces higher dust temperatures). This makes the upper
($T_w=40$ K) curves more plausible although their shape is still
sensitive to the mass fraction and temperature of the cold dust
(colder dust or more cold dust will cause the indicator to turn over
earlier and flatten more), and is also dependent on the value of
$\beta$. If in reality, $\beta$ lies somewhere between $1.3-2$, then
this will make the `grey' models more like the median one at low
redshifts. Figure~\ref{greyF} suggests that the uncertainties in
whether there is actually a single dust temperature, or in the true
value of $\beta$, do not add much to the uncertainty we have derived
from the single-temperature fits. Currently, it is not possible to be
more specific about the nature of any cold dust components as there
are only a handful of local galaxies with enough FIR and submm fluxes
to make a decomposition of the SED feasible. Our knowledge of dust
properties in local galaxies should improve in the near future (Dunne
et al. 2000).}
\item{Contributions to the radio flux by AGN will produce misleading
redshift estimates if the object is not recognised as such. Since the
fraction of deep SCUBA sources harbouring radio-loud AGN is still not
well determined, the {\em estimator must be treated as a statistical tool}
rather than a redshift indicator for individual objects. The redshifts
given by the upper $1 \sigma$ curve should be treated as a robust
lower limit.}
\item{A variation in the FIR-radio correlation with redshift is quite 
possible if either magnetic fields or dust properties were different
in the past, although currently this is difficult to test. Condon
(1992) does however point out that the FIR-radio relation in the local 
universe holds over a large range of magnetic field strengths. The
tendency of the galaxies observed so far in both the radio and
submillimetre at high redshifts, to lie in agreement with the
estimator (Fig.~\ref{car2F}) suggests that there has been no dramatic
evolution in the underlying FIR-radio relation, although more data
is needed to fully investigate this.}
\end{enumerate} 

\section{Conclusions}

We have used the submm data from a large, complete sample of local
{\em IRAS\/} galaxies, along with complementary radio data, to define
a median redshift estimator using the change in spectral slope between
the submm and the radio at $\sim 3$mm. The scatter in the data has
been used to provide $1 \sigma$ uncertainties on the relationship. The
limited number of submm sources from the deep surveys with both radio
fluxes and spectroscopic redshifts generally agree with the estimator
(within the uncertainties), except for one object which is known to be
an AGN (Fig~\ref{car2F}). {\em The estimator is useful in a
statistical sense rather than for predicting the redshifts of
individual objects\/} as there are many uncertainties, especially the
possible contribution to the radio flux by AGN which would lower the
data relative to the models thus leading to an under-estimated
redshift. Our redshift indicator differs from the recent one of
Carilli \& Yun (2000) in terms of shape and normalisation. The
difference is significant, particularly at redshifts $>2$
(Fig~\ref{carF}). The discrepancy can be attributed primarily to the
larger sample size used in this work and also, in part, to the
under-estimation of submm fluxes used in CY 2000. An
erratum to CY 2000 has since been published which
produces a revised redshift estimator using the corrected Lisenfeld et
al. (2000) submm fluxes. This removes $\sim 30$ per cent of the
discrepancy between the estimators, and we postulate that the
remaining difference is a result of sample selection (the Lisenfeld
sample is not a complete sample and contains a higher proportion of
radio bright objects compared to SLUGS), and the smaller sample size (17
objects compared to 104 in SLUGS).

The submm-radio spectral index decreases as radio and FIR luminosity
increase, but shows no strong trend with 850$\mu$m luminosity
(Fig.~\ref{alpha_lumF}). The correlations are believed to be due to the
sensitivity of the non-thermal radio and thermal FIR emission to recent star
formation, while a significant fraction of the 850$\mu$m flux may be
due to colder dust, and not linked to the star formation
rate. The absence of a correlation between $\alpha^{850}_{1.4}$ and
850$\mu$m luminosity means that any differences in $L_{850}$ between
SLUGS and the SCUBA-selected galaxies will not lead to biased redshift 
estimates. Importantly, we also find no significant evidence for a
systematic variation of $\alpha^{850}_{1.4}$ with dust temperature
(Fig.~\ref{alpha_tempF}) implying that the redshift-temperature
degeneracy hypothesised by Blain (1999) does not play a dominant part in the
scatter of the $\alpha^{850}_{1.4}$ -- redshift relation.

We have investigated the impact of neglecting a possible colder dust
component in the SEDs of the local galaxies (Fig~\ref{greyF}). While
sensitive to the parameters assumed, in general the differences are
not very significant for most plausible models and the agreement is
still within the $1 \sigma$ uncertainties at lower redshifts
($<2$). If cold components are present but unaccounted for, using the
median line in Fig~\ref{carF} to estimate redshifts will most likely
lead to an over-estimate of $z$.
\vspace{0.2cm}

We would like to thank Chris Carilli and Ute Lisenfeld for useful
discussions. The support of PPARC is gratefully aknowledged by
L. Dunne, S. Eales and D. Clements.

\end{document}